# Robust Functional Data Analysis for Discretely Observed Data

Lingxuan Shao

*Department of Statistics and Data Science, School of Management, Fudan University, Shanghai, China*

E-mail: shao_lingxuan@fudan.edu.cn

Fang Yao†

*Center for Statistical Science, Department of Probability and Statistics, School of Mathematical Sciences, Peking University, Beijing, China*

E-mail: fyao@math.pku.edu.cn

**Summary**. This paper examines robust functional data analysis for discretely observed data, where the underlying process encompasses various distributions, such as heavy tail, skewness, or contaminations. We propose a unified robust concept of functional mean, covariance, and principal component analysis, while existing methods and definitions often differ from one another or only address fully observed functions (the "ideal" case). Specifically, the robust functional mean can deviate from its non-robust counterpart and is estimated using robust local linear regression. Moreover, we define a new robust functional covariance that shares useful properties with the classic version. Importantly, this covariance yields the robust version of Karhunen–Loève decomposition and corresponding principal components beneficial for dimension reduction. The theoretical results of the robust functional mean, covariance, and eigenfunction estimates, based on pooling discretely observed data (ranging from sparse to dense), are established and align with their non-robust counterparts. The newly-proposed perturbation bounds for estimated eigenfunctions, with indexes allowed to grow with sample size, lay the foundation for further modeling based on robust functional principal component analysis.

## 1. Introduction

Functional data analysis has emerged as a valuable tool for modeling data with repeated functional observations and has been extensively studied over the past decades. For instance, the monographs Ramsay and Silverman (2005); Hsing and Eubank (2015); Kokoszka and Reimherr (2017), as well as the survey paper Wang et al. (2016), provide a comprehensive overview of functional data analysis. In practice, the underlying process may be drawn from various populations, such as those with heavy tails, skewness, or mixed types. Robust analysis of functional data has attracted substantial interest in many real-world applications (Lugosi and Mendelson, 2019; Boente and Salibián-Barrera, 2021), which is the main focus of this paper. Furthermore, such data are often observed intermittently, particularly with sparseness, posing great challenges to the development of methodology and theory. It is crucial to distinguish between fully and discretely observed functional data; the former represents an "ideal" scenario in which underlying random trajectories are treated as observed, while the latter accounts for discretization and noise contamination, reflecting more realistic practical situations. In what follows, we consider an irregular sampling scheme that typically assumes randomly observed times, also referred to as the "independent design" in Cai et al. (2011), which is convenient for analyzing the challenging sparse cases.

†Fang Yao is the corresponding author. His research is partially supported by the National Natural Science Foundation of China (No. 12292981, 11931001, 11871080), the National Key R&D Program of China (No. 2022YFA1003801, 2020YFE0204200), the LMAM and the LMEQF.

The research on robustness typically consists of two types, one of which concerns outlier-detection serving as data pre-processing procedures, and the other deals with robust modeling considering data sampled from skewed or heavy-tailed populations. The former aims to exclude outliers by identifying potential atypical observations and has been developed only for fully observed functional data (Ren et al., 2017; Nieto-Reyes and Battey, 2016; Sun and Genton, 2011; Dai et al., 2020). The robust modeling methods, utilizing a robust loss function, can be traced back to Huber (Huber, 1964), and classical nonparametric approaches include robust local polynomial regression (Fan and Truong, 1994; Jiang and Mack, 2001), local composite quantile regression (Kai et al., 2010), among others. For functional data, most of the existing literature focuses on the fully observed case (Sinova et al., 2018; Cardot and Godichon-Baggioni, 2017; Dubey and Müller, 2021; Bali et al., 2011), with some extensions to the discretely observed case (Lima et al., 2018; Kalogridis and Van Aelst, 2022). However, these works did not provide a unified framework for robust mean and covariance modeling that is easy to compute, satisfies some desirable properties, and naturally leads to a robust principal component decomposition related to the underlying population. Therefore, our objective is to develop such a framework, applicable for robust mean, covariance, and eigen-analysis of discretely observed data.

A major challenge in robust functional data analysis arises from the absence of a consensus on defining the mean and covariance structures. We initiate our discussion with **robust mean estimation** and contemplate the general scenario where the robust mean, denoted by $\mu_r = \arg\min_\beta \rho(X-\beta)$ for a robust loss $\rho$, may not coincide with the classical (non-robust) mean, denoted by $\mu_c = \mathsf{E}(X)$. This discrepancy is particularly noticeable for asymmetric or skewed populations commonly encountered in practice. The robust mean, $\mu_r$, akin to the median or Huber mean, characterizes the location of the population in a more stable manner compared to the classical mean, $\mu_c$. Nevertheless, a majority of existing research, including Lima et al. (2018); Kalogridis and Van Aelst (2022), has made the restrictive assumption of $\mu_r = \mu_c$, which lacks generality. Although Sinova et al. (2018) discerned the robust functional mean from its classical counterpart, their analysis solely considered fully observed functions. In this paper, we utilize a pooling estimation strategy for discretely observed functional data via robust local linear regression and establish its asymptotic properties, which align with those under classical squared loss (Zhang and Wang, 2016).

The **robust covariance function** has been defined in various ways for functional data, with most of the focus on fully observed data. For instance, Cardot and Godichon-Baggioni (2017) defined the median covariance as a generalization of the median for random variables in a Hilbert space. Dubey and Müller (2021) considered metric-space-valued random processes and defined the robust covariance as the covariance of the distance process, which is, in fact, a fourth moment and only contains distance-related information. Zhong et al. (2022) considered discretely observed data and introduced Kendall's $\tau$ covariance function with the aid of the independent copy technique such that the eigenfunctions are the same as those of the classic covariance. It is worth noting that these robust covariances are materially different from each other and cannot be consolidated into a unified framework serving as a natural generalization of the classical covariance.

To this end, we define the robust functional covariance which shares useful properties with its classic counterpart and incorporates the robust mean as a whole. For instance, this robust covariance of two random variables is equal to zero when they are independent and preserves the sign/direction information of the random variables under a symmetric loss function. More importantly, the eigen-decomposition of this robust covariance naturally leads to the Karhunen–Loève decomposition of the underlying population, and subsequently provides a practical framework to study the **robust functional principal components analysis** (robust FPCA) that is essential for further modeling, such as functional linear regression and so on. From a computational perspective, our robust covariance function and its corresponding estimate admit analytical expressions when provided with mean estimate, thereby circumventing the need for numerical optimization during covariance estimation and facilitating fast computations. By contrast, existing research (Cardot and Godichon-Baggioni, 2017; Dubey and Müller, 2021; Bali et al., 2011; Zhong et al., 2022) on robust functional covariance and eigen-decomposition cannot achieve all these advantages

On the theoretical side, the convergence property of the proposed covariance estimate is established with the same rate and phase transition phenomenon depending on the sampling frequency relative to the sample size, as observed in the classic scenario (Zhang and Wang, 2016, Theorem 4.2). Furthermore, we adopt the newly developed perturbation techniques in Zhou et al. (2022) for the eigen-decomposition of the robust covariance estimation obtained by pooling discretely observed data using the kernel method. This takes into account the unknown mean, which is technically nontrivial to tackle the robust mean estimate that has no analytic expression. Specifically, our analysis reveals the connection among the number of components $k_0$ that is allowed to diverge, the sample size $n$, and the sampling frequency $m$, which illustrates a phase transition phenomenon different from the mean/covariance estimation and reflects the elevated difficulty in estimating an increasing number of robust eigenfunctions. This provides a foundation for exploring further models based on robust FPCA.

The paper is organized as follows. Section 2 introduces the robust functional model with mean, newly defined covariance, and eigenfunctions. Estimation procedures and theoretical results are presented in Sections 3 and 4, respectively. Numerical studies, including simulations in Section 5 and an application to Alzheimer's data in Section 6, are discussed. Proofs of theorems and ancillary lemmas, along with additional data analysis results, are deferred to the online Supplementary Material for space economy.

## 2.  Robust Functional Model

### 2.1.  Robust Mean and Covariance

In this section, we first propose the robust mean and covariance for a random variable $X : \Omega \to \mathbb{R}$, as the groundwork for the functional versions. The robust mean for this random variable $X$ is widely acknowledged as

$$\mu_r = \arg\min_{\beta \in \mathbb{R}} \mathsf{E}\, \rho(X - \beta), \tag{1}$$

where the robust loss function $\rho$ is detailed in Section 2.3. This robust mean, denoted as $\mu_r$, is equivalently the solution of $\mathsf{E}\, \psi(X - \beta) = 0$ with $\psi = \rho'$ under mild conditions. In comparison to the classic version $\mathsf{E}\,(X - \mu_c) = 0$, the robust deviation $(X - \mu_r)$ has an expectation of zero after rescaling by $\psi$, meaning $\mathsf{E}\, \psi(X - \mu_r) = 0$. Fig. 1 illustrates two perspectives of defining the robust mean.

$$\mathsf{E}\,(X - \mu_c) = 0 \quad \xrightarrow{\text{replace the identity function by a rescaling function } \psi} \quad \mathsf{E}\, \psi(X - \mu_r) = 0$$

$$\mu_c = \arg\min_{\beta \in \mathbb{R}} \mathsf{E}\,(X - \beta)^2 \quad \xrightarrow{\text{replace square function by a robust loss function } \rho} \quad \mu_r = \arg\min_{\beta \in \mathbb{R}} \mathsf{E}\, \rho(X - \beta)$$

**Fig. 1.**  The robust mean is an extension of classic mean from two perspectives.

In this paper we take the general perspective that $\mu_c$ and $\mu_r$ are potentially different parameters, while Kalogridis and Van Aelst (2022); Lima et al. (2018) presented the assumption $\mu_c = \mu_r$ implicitly by imposing the condition that $\epsilon_i(T_{ij}) = X_i(T_{ij}) - \mu_c(T_{ij})$ satisfies $\mathsf{E}\, \rho'(\epsilon_{ij}) = 0$, which implies that $\mu_r(T_{ij}) = \arg\min_{\beta \in \mathbb{R}} \mathsf{E}\, \rho(X_{ij} - \beta)$ equals $\mu_c(T_{ij})$ and thus $\mu_c = \mu_r$.

The definition of robust covariance has different forms in existing literature (Cardot and Godichon-Baggioni, 2017; Dubey and Müller, 2021; Zhong et al., 2022; Minsker, 2018; Ke et al., 2019). We aim to propose a new definition of robust covariance $C_r$ satisfying the following properties.

(1) The robust covariance $C_r$ is symmetric and coincides with the classic covariance under the square loss function $\rho = (\cdot)^2$ or identity rescaling function $\psi = \mathbf{I}$.

(2) The robust covariance $C_r$ integrates the robust mean $\mu_r$, but not the classic mean $\mu_c$.

(3) The robust covariance $C_r$ of $X_1, X_2$ equals zero if $X_1, X_2$ are independent.

(4) The robust covariance $C_r$ preserves sign/direction when $\rho$ is symmetric, meaning

$$C_r(X_1, X_2) = -C_r(-X_1, X_2) = -C_r(X_1, -X_2).$$

Property (1) is a natural requirement in robust analysis, and (2) presents a unified robust framework for both mean and covariance. Property (3) puts a constraint on the "robust linear dependence" of the robust covariance in the sense that the random variables share no such dependence if they are independent, while (4) ensures that the "robust linear dependence" is capable of identifying the negative sign. Moreover, Properties (3) and (4) serve as the rudiments for constructing the associated eigen-decomposition, which will be discussed in the subsequent section. Taking into account the above considerations, we propose the *robust covariance* by substituting $X_i - \mu_c(X_i)$ in the classic version with the rescaling form $\psi(X_i - \mu_r(X_i))$ for $i = 1, 2$,

$$C_r(X_1, X_2) = E\left\{\psi\left(X_1 - \mu_r(X_1)\right) \times \psi\left(X_2 - \mu_r(X_2)\right)\right\} \quad (2)$$

where each random variable is centered by its robust mean and rescaled by the rescaling function $\psi$. As demonstrated in the subsequent lemma, this robust covariance indeed meets the aforementioned properties, aligning with our expectations.

‡The notation $\rho'$ refers to the derivative of $\rho$.

Lemma 2.1. *The robust covariance, as defined in Equation* (2), *fulfills the Properties* (1), (2), (3) *and* (4).

To the best of our knowledge, the existing literature on robust covariance definitions fails to fulfill the aforementioned properties. For instance, Zhong et al. (2022) introduced the Kendall's $\tau$ function

$$K(s, t) = E\left[\frac{\{X(s) - \tilde{X}(s)\}\{X(t) - \tilde{X}(t)\}}{\int |X(u) - \tilde{X}(u)|^2 du}\right],$$

where $\tilde{X}$ is an independent copy of $X$. This Kendall's $\tau$ function can not be reduced to classic form and thus can not realize Property (1). Cardot and Godichon-Baggioni (2017) defined the median covariance as

$$C_r(s, t) = \arg\min_{\beta \in L(H;H)} E\left\{\left\|X(s)X(t)^\top - \beta\right\| - \left\|X(s)X(t)^\top\right\|\right\}$$

for zero-median process $X$ and Hilbert space $H$, which does not fulfill Property (3). Our robust covariance definition rescales $\{X_1 - \mu_r(X_1)\}$ and $\{X_2 - \mu_r(X_2)\}$, respectively, before multiplication, while Minsker (2018); Ke et al. (2019) rescale after multiplication, resulting in the form of $\psi\{(X_1 - \mu_r(X_1))(X_2 - \mu_r(X_2))\}$. This difference allows our robust covariance to satisfy Property (3), while Minsker (2018); Ke et al. (2019) do not. Dubey and Müller (2021) considered the metric-space-valued random process and defined

$$C_r(s, t) = E(V(s)V(t)) - E\{V(s)\}E\{V(t)\}$$

with distance process $V(t) = d^2(X(t), \mu_r(t))$, which violates Property (4).

## 2.2. Robust Functional Model

Functional data analysis involves the study of samples drawn from a random process $X : \Omega \times T \to R$, where $T$ is a closed time interval in $R$. Building upon the previous discussion on equations (1) and (2), the robust functional mean and covariance can be defined for $s, t \in T$ as

$$\mu_r(t) = \arg\min_{\beta \in R} E\,\rho\left(X(t) - \beta\right) = \text{the solution of } E\,\psi\left(X(t) - \beta\right) = 0; \quad (3)$$

$$C_r(s, t) = E\left\{\psi\left(X(s) - \mu_r(s)\right) \times \psi\left(X(t) - \mu_r(t)\right)\right\}. \quad (4)$$

Note that the robust functional covariance specifically targets the *rescaled process* $Z(t) = \psi(X(t) - \mu_r(t))$, and can be represented as $C_r(s, t) = E\{Z(s)Z(t)\}$. The primary advantage of this robust covariance definition is its innate ability to result in the following robust eigen-decomposition

$$C_r(s, t) = \sum_{k=1}^{\infty} \lambda_k \varphi_k(s)\varphi_k(t) \tag{5}$$

where $\{\lambda_k\}_{k=1}^{\infty}$ are the *robust eigenvalues* satisfying $\lambda_1 \geq \lambda_2 \geq \ldots$ and $\{\varphi_k\}_{k=1}^{\infty}$ are the corresponding orthonormal *robust eigenfunctions*. Then the robust Karhunen–Loève decomposition of the population $X$ is given by

$$\psi(X(t) - \mu_r(t)) = Z(t) = \sum_{k=1}^{\infty} \xi_k \varphi_k(t), \tag{6}$$

where the $k$th *robust functional principal component score* $\xi_k$ is computed as $\xi_k = \int Z(t)\varphi_k(t)dt$ (Hsing and Eubank, 2015, Theorem 7.3.5). The robust Karhunen–Loève decomposition demonstrates that there exists a significant connection between the robust eigenfunctions and the population through the rescaling function $\psi$. This connection provides a comprehensive framework for robust feature extraction. For instance, the robust principal components or the robust $K$-dimension projection is

$$X^{(K)}(t) = \mu_r(t) + \psi^{-1}\left(\sum_{k=1}^{K} \xi_k \varphi_k(t)\right) \tag{7}$$

if the rescaling function $\psi$ is strictly monotonic, which is widely satisfied for globally smooth loss function discussed in the next Section, or may be broadly defined as $\psi^{-1}(t) = \inf\{s : \psi(s) \geq t\}$ or $\psi^{-1}(t) = \sup\{s : \psi(s) \leq t\}$, whichever is appropriate. Consequently, additional statistical models that rely on dimension reduction methods, including functional linear regression (Yao et al., 2005; Hall and Horowitz, 2007; Zhou et al., 2022), can be expanded using this robust dimension reduction approach as shown in Equation (7). However, such a valuable relationship was not presented in the work of Bali et al. (2011).

### *2.3. Robust Loss Function*

In our framework, the robust loss function is expected to be sufficiently regular due to the presence of the rescaling function $\psi = \rho'$, as well as the third-order differentiability required by Assumption 4.1 (2), as discussed in Theory Section 4. Nevertheless, several frequently encountered robust loss functions, such as the absolute loss and Huber loss, do not satisfy these requirements. To address this issue, we introduce two approaches to smooth the loss functions, which are widely adopted in the literature: *local smooth* and *global smooth* methods. Both smoothing techniques are compatible with our framework, and we implement them in our numerical studies.

*(1)* **Locally smooth loss.** For a non-smooth robust loss function $\rho$ with a discontinuous point $x_0$, we replace the segment $\rho|_{[x_0 - \kappa, x_0 + \kappa]}$ by a smooth approximation. Specifically, we consider a third-order-differentiable modification of the absolute loss function $\rho(x) = |x|$ with hyper-parameter $\kappa$

$$\rho_1(x) = |x|I_{|x|>\kappa} + \frac{1}{8\kappa}(-x^4 + 6\kappa^2 x^2 + 3\kappa^4)I_{|x|\leq\kappa}, \tag{8}$$

shown in the left panel of Fig. 2. The selection of the hyper-parameter $\kappa$ will be discussed in Section 3.4.

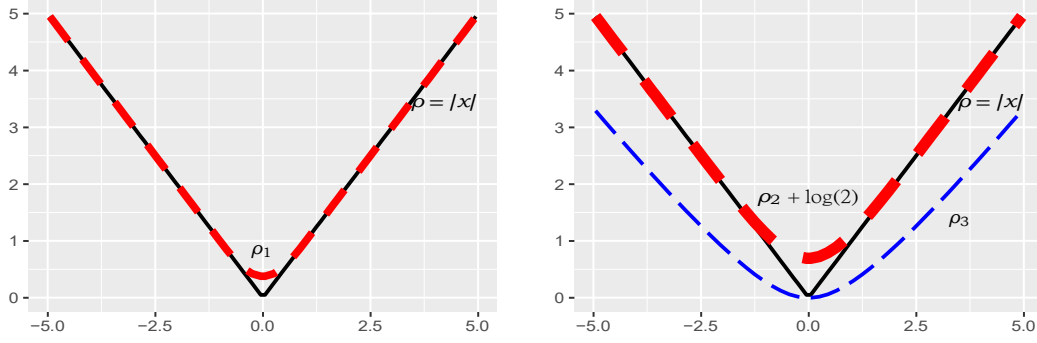

**Fig. 2.** The left panel depicts the absolute loss function $\rho(x) = |x|$ (black solid line) and the locally smooth robust loss function $\rho_1$ with $\kappa = 1$ (red dotted line), as described in (8). The right panel illustrates the globally smooth loss functions, where the black solid line represents the absolute loss function $\rho(x) = |x|$ and the red dotted and blue dashed lines correspond to the global modifications $\rho_2 + \log(2)$ and $\rho_3$, as defined in (9), respectively.

(2) **Globally smooth loss.** Alternatively, a non-smooth robust loss function can be globally transformed into a smooth one exhibiting a similar pattern. For instance, the discontinuous derivative $\rho'(x) = I_{x>0} - I_{x<0}$ of the absolute loss $\rho(x) = |x|$ can be altered to achieve identical limiting behavior as $x \to \pm\infty$ and to maintain smoothness around the original point $x = 0$. Two such transformations include $(e^x - e^{-x})/(e^x + e^{-x})$ and $2\pi^{-1} \arctan(x)$, which means that the globally modified smooth robust loss functions corresponding to the absolute loss $\rho(x) = |x|$ are the integral functions of $(e^x - e^{-x})/(e^x + e^{-x})$ and $2\pi^{-1}\arctan(x)$,

$$\rho_2(x) = \log(\cosh(x)), \quad \rho_3(x) = 2\pi^{-1} x \arctan(x) - \pi^{-1} \ln(1 + x^2). \tag{9}$$

Using these constructions, both $\rho_2$ and $\rho_3$ are globally smooth, and exhibit limiting behaviors closely resembling the absolute loss, as evidenced by $\rho_2 \sim |x| - \log(2)$ and $\rho_3 \sim |x| - 2\pi^{-1} \ln|x|$. These relationships are illustrated in the right panel of Fig. 2.

## 3. Estimation

In this section, we aim to estimate the robust mean, covariance, and eigenfunctions from discrete observations. Specifically, we consider the discrete observations $\{X_{ij} = X_i(T_{ij}) : 1 \leq i \leq n, 1 \leq j \leq m_i\}$, which are sampled from the independent and identically distributed realizations $\{X_i : 1 \leq i \leq n\}$ drawn from the random process $X$. The sample size, denoted by $n$, represents the number of subjects (curves), while $m_i$ denotes the number of noisy measurements per subject, referred to as the sampling frequency. Frequently, it is assumed that $m_i = m \asymp n^\alpha$§, where $\alpha$ is the sampling rate. We also assume that the time points $\{T_{ij} : 1 \leq i \leq n, 1 \leq j \leq m_i\}$ are uniformly distributed on $\mathcal{T} = [0, 1]$ for a clearer exposition; extending this to more general cases is technically straightforward (Zhang and Wang, 2016).

---

§The notation $a_n \lesssim b_n$ signifies that $\frac{a_n}{b_n} = O(1)$ for large $n$, while $a_n \asymp b_n$ implies both $a_n \lesssim b_n$ and $b_n \gtrsim a_n$.

Assumption 3.1. *The observed time points $T_{ij} : 1 \leq i \leq n, 1 \leq j \leq m_i$ represent independent and uniformly distributed samples, supported on a compact interval* $[0, 1]$. *These time points are also independent of the random process $X$.*

### 3.1. Robust Mean Estimation

We modify the local linear regression approach to estimate the robust mean function by utilizing a robust loss function. The classical local linear estimate of the functional mean, derived from discrete observations, is expressed as

$$\hat{\mu}_c(t) = \arg_{\beta_0} \min_{\beta_0, \beta_1} \sum_{i=1}^{n} \gamma_i \sum_{j=1}^{m} K_{h_\mu}(T_{ij} - t) \left\{ X_{ij} - \beta_0 - \beta_1(T_{ij} - t) \right\}^2 \quad (10)$$

where $\{\gamma_i\}_{1 \leq i \leq n}$ can be subject-specific weights and are set as $\gamma_i = (nm)^{-1}$ for simlicity (Li and Hsing, 2010; Zhang and Wang, 2016), and $K_{h_\mu}$ is a kernel function $K_{h_\mu}(\cdot) = h_\mu^{-1} K(\cdot/h_\mu)$. In order to provide motivation to propose the robust local estimate, we address that the classical local linear estimate $\hat{\mu}_c$ can be regarded as being derived through the following two steps: (1) incorporating a kernel with bandwidth $h_\mu$ and (2) replacing the population distribution with its empirical counterpart, akin to the approach adopted in Petersen and Müller (2019).

$$\mu_c(t) \xrightarrow{\text{incorporating a kernel}} \tilde{\mu}_c(t) \xrightarrow{\text{adopt empirical distribution}} \hat{\mu}_c(t)$$

The middle item, $\tilde{\mu}_c$, denoted as the localized mean, is obtained by reverting the empirical distribution in $\hat{\mu}_c(t)$ back to its population version. A straightforward computation (Lemma S.4.1 in the Supplementary Material) results in the subsequent expression

$$\tilde{\mu}_c(t) = \arg_{\beta_0} \left\{ \min_{\beta_0, \beta_1} \mathsf{E} \left[ K_{h_\mu}(T - t) \left\{ X - \beta_0 - \beta_1(T - t) \right\}^2 \right] \right\}$$
$$= \arg \min_{\beta} \mathsf{E} \left\{ \omega(T, t, h_\mu)(X - \beta)^2 \right\}$$

where the weight function, referred to as the *local extension multiplier*, is $\omega(T, t, h_\mu) = \sigma_0^{-2} K_{h_\mu}(T - t)\{u_2 - u_1(T - t)\}$ with $u_j = \mathsf{E}\{K_{h_\mu}(T - t) \times (T - t)^j\}$ and $\sigma_0^2 = u_0 u_2 - u_1^2$. The localized mean, denoted by $\tilde{\mu}_c(t)$, is characterized by this local extension multiplier $\omega(T, t, h_\mu)$ encompassing the support set around the time point $t$ and fulfilling the condition $\mathsf{E}\,\omega(T, t, h_\mu) = 1$. Thus, $\tilde{\mu}_c(t)$ can be perceived as a local extension of the classic mean, $\mu_c(t)$, while the classical local linear estimate, $\hat{\mu}_c(t)$, serves as the empirical approximation of $\tilde{\mu}_c(t)$. Motivated by this insight, we seek to estimate $\mu_r(t)$ within a robust framework, as expounded below.

$$\mu_r(t) \xrightarrow{\text{incorporating a kernel}} \tilde{\mu}_r(t) \xrightarrow{\text{adopt empirical distribution}} \hat{\mu}_r(t)$$

Analogous to $\tilde{\mu}_c$, the localized robust mean $\tilde{\mu}_r$ is formulated by incorporating the local extension multiplier $\omega(T, t, h_\mu)$.

$$\tilde{\mu}_r(t) = \arg\min_\beta \mathbb{E}\left\{\omega(T, t, h_\mu)\rho(X - \beta)\right\}.$$

Subsequently, the robust local linear estimate of the functional mean $\mu_r$ in (3) is acquired by adopting the empirical distribution of $(X, T)$ as

$$\hat{\mu}_r(t) = \arg\min_\beta \sum_{i=1}^n \gamma_i \sum_{j=1}^m \hat{\omega}(T_{ij}, t, h_\mu)\rho\left(X_{ij} - \beta\right) \tag{11}$$

where $\{\gamma_i\}_{1\leq i \leq n} = (nm)^{-1}$, $\hat{\omega}(T_{ij}, t, h_\mu) = \hat{\sigma}_0^{-2} K_{h_\mu}(T_{ij} - t)[\hat{u}_2 - \hat{u}_1(T_{ij} - t)]$ with $\hat{u}_l = \sum_i \gamma_i \sum_j K_{h_\mu}(T_{ij} - t) \times (T_{ij} - t)^l$ for $l = 0, 1, 2$ and $\hat{\sigma}_0^2 = \hat{u}_0 \hat{u}_2 - \hat{u}_1^2$. The bandwidth $h_\mu$ is selected by a $K$-fold cross validation by leaving out whole subjects. That is to randomly and equally split $n$ subjects into $K$ subsets $\{I_1, \ldots, I_K\}$, and use each subset in turn for validation to minimize the empirical robust error $\sum_{k=1}^K \sum_{i \in I_k} \gamma_i \sum_j \rho\left(X_{ij} - \hat{\mu}_r^{(-k)}(T_{ij})\right)$, where $\hat{\mu}_r^{(-k)}$ is the robust mean estimate using the training set after excluding the subjects in $I_k$.

Comparatively, a direct robust local linear estimate

$$\hat{\mu}_{r,1}(t) = \arg_{\beta_0}\min_{\beta_0, \beta_1} \sum_{i=1}^n \gamma_i \sum_{j=1}^{m_i} K_h(T_{ij} - t)\rho\left(X_{ij} - \beta_0 - \beta_1(T_{ij} - t)\right)$$

can be obtained via substituting the square function with a robust loss function in (10). This method involves a bivariate numerical minimization over two unknown parameters, $\beta_0$ and $\beta_1$, resulting in a more computationally intensive and technically complex process than our proposed estimate, $\hat{\mu}_r$, which only requires univariate numerical minimization with respect to a single unknown parameter, $\beta$.

Alternatively, a robust local constant estimate can be given by

$$\hat{\mu}_{r,2}(t) = \arg\min_\beta \sum_i \gamma_i \sum_j K_h(T_{ij} - t)\rho(X_{ij} - \beta),$$

which includes only one unknown parameter but exhibits sub-optimal convergence performance near the boundaries. This issue is already evident in the classical case, as discussed by Hsing and Eubank (2015).

### 3.2. Robust Covariance Estimation

Given the robust functional covariance in (4), and the robust mean estimate in (11), we begin by formulating the robust raw covariance observations as

$$\hat{C}_{i,j_1,j_2} = \psi\left(X_{ij_1} - \hat{\mu}_r(T_{ij_1})\right) \times \psi\left(X_{ij_2} - \hat{\mu}_r(T_{ij_2})\right).$$

The local linear regression is applied to the data $\{T_{ij_1}, T_{ij_2}, \hat{C}_{i,j_1,j_2} : 1 \leq j_1 \neq j_2 \leq m, i = 1, \ldots, n\}$ and yields

$$(\hat{\beta}_0, \hat{\beta}_1, \hat{\beta}_2) = \arg\min_{\beta_0, \beta_1, \beta_2 \in \mathbb{R}} \sum_{i=1}^n v_i \sum_{j_1 \neq j_2, 1 \leq j_1, j_2 \leq m} K_{h_C}(T_{ij_1} - s) K_{h_C}(T_{ij_2} - t)$$
$$\times \left(\hat{C}_{i,j_1,j_2} - \beta_0 - \beta_1(T_{ij_1} - s) - \beta_2(T_{ij_2} - t)\right)^2,$$

where $\{v_i\}_{1\leq i\leq n}$ can also be subject-specific weights and are set as $v_i = \{nm(m-1)\}^{-1}$ for simplicity, and $K_{h_C}$ is a kernel function $K_{h_C}(\cdot) = h_C^{-1}K(\cdot/h_C)$. Here we retain the square loss for estimating the robust covariance, thereby avoiding the introduction of a new robust loss function. This is because the rescaling function $\psi$ utilized in the calculation of the robust raw covariances already provides the desired level of robustness. Consequently, the minimizer $\hat{\beta}_0$ can be expressed explicitly as $\hat{\beta}_0 = \hat{w}_1\hat{R}_{00} + \hat{w}_2\hat{R}_{10} + \hat{w}_3\hat{R}_{01}$, where $\hat{R}_{ab}$ and $\hat{w}_i$ are

$$\hat{R}_{ab} = \sum_i v_i \sum_{j_1\neq j_2}(T_{ij_1}-s)^a(T_{ij_2}-t)^b K_{h_C}(T_{ij_1}-s)K_{h_C}(T_{ij_2}-t)\hat{C}_{i,j_1,j_2},$$

$$\hat{S}_{ab} = \sum_i v_i \sum_{j_1\neq j_2}(T_{ij_1}-s)^a(T_{ij^2}-t)^b K_{h_C}(T_{ij_1}-s)K_{h_C}(T_{ij_2}-t),$$

$$\hat{W}_1 = (\hat{S}_{20}\hat{S}_{02} - \hat{S}_1^2), \quad \hat{W}_2 = -(\hat{S}_{10}\hat{S}_{02} - \hat{S}_{01}\hat{S}_{11}), \quad \hat{W}_3 = (\hat{S}_{10}\hat{S}_{11} - \hat{S}_{01}\hat{S}_{20}),$$

$$\hat{w}_i = \hat{W}_i(\hat{W}_1\hat{S}_{00} + \hat{W}_2\hat{S}_{10} + \hat{W}_3\hat{S}_{01})^{-1}, \text{ for } 1 \leq i \leq 3.$$

We observe that $\hat{S}_{ab}$ depends solely on $\{T_{ij}\}$, which are assumed to be uniformly distributed. This implies that the targeted values $S_{ab} = \mathbb{E}\hat{S}_{ab}$ can be computed analytically or approximated by Monte Carlo averages over independently simulated times. Consequently, we simplify the estimation of the robust functional covariance $C_r(s, t)$ by replacing $\hat{S}_{ab}$ with such Monte Carlo estimates $\tilde{S}_{ab}$, which further facilitates the theoretical analysis.

$$\hat{C}_r(s, t) = \tilde{w}_1\hat{R}_{00} + \tilde{w}_2\hat{R}_{10} + \tilde{w}_3\hat{R}_{01}, \quad (12)$$

where $\tilde{w}_i = \tilde{W}_i(\tilde{W}_1\tilde{S}_{00} + \tilde{W}_2\tilde{S}_{10} + \tilde{W}_3\tilde{S}_{01})^{-1}$, $1 \leq i \leq 3$, $\tilde{W}_1 = (\tilde{S}_{20}\tilde{S}_{02} - \tilde{S}_1^2)$, $\tilde{W}_2 = -(\tilde{S}_{10}\tilde{S}_{02} - \tilde{S}_{01}\tilde{S}_{11})$, $\tilde{W}_3 = (\tilde{S}_{10}\tilde{S}_{11} - \tilde{S}_{01}\tilde{S}_{20})$. The bandwidth $h_C$ is selected in a similar way to $h_\mu$ by minimizing a $K$-fold cross-validation using the empirical square error over the validation set, i.e. $\sum_{k=1}^K \sum_{i\in I_k} v_i \sum_{j_1\neq j_2}\{\hat{C}_{i,j_1,j_2} - \hat{C}_r^{(-k)}(T_{ij_1}, T_{ij_2})\}^2$, where $\hat{C}_r^{(-k)}$ is the robust covariance estimate using the training set after excluding the subjects in $I_k$.

### 3.3. Robust Principal Component Analysis

The eigen-decomposition of the covariance estimate in (12) results in

$$\hat{C}_r(s, t) = \sum_{k=1}^\infty \hat{\lambda}_k \hat{\varphi}_k(s)\hat{\varphi}_k(t) \quad (13)$$

where $\{\hat{\varphi}_k\}_{1\leq k\leq \infty}$ are pairwise orthonormal in the sense that $\int_T \hat{\varphi}_k(t)\hat{\varphi}_l(t)dt = I_{k=l}$. The robust eigenfunction $\lambda_k$ and eigenvalue $\varphi_k$ in (5) are estimated by $\hat{\lambda}_k$ and $\hat{\varphi}_k$, respectively, where the sign of the eigenfunction estimate is chosen to satisfy $\int_T \hat{\varphi}_k(t)\varphi_k(t)dt > 0$.

In estimating the score, we utilize the sampling average

$$\hat{\xi}_{ik} = \frac{1}{m} \sum_{j=1}^m \psi'\left(X_{ij} - \hat{\mu}_r(T_{ij})\right)\hat{\varphi}_k(T_{ij}).$$

This approach has shown satisfactory numerical performance compared with the PACE method (Yao et al., 2005) even with sparse schemes, as evidenced by simulations provided in the Supplementary Material and related research (Zhou et al., 2022). It is worth noting, however, that the PACE method proposed by Yao et al. (2005) may not be appropriate for our robust framework, due to the fact that both the underlying population $X(t)$ and the rescaled process $Z(t)$ frequently deviate from the Gaussian assumption that is required in PACE method.

Remark 3.1. *In order to address the technical challenges arising from the dependence of robust eigenfunction estimate on the mean estimate, which does not possess an analytic form as opposed to the classical case, we can employ a straightforward sample-splitting approach. Specifically, the sample can be evenly partitioned into two subsets: $\{(X_{ij}, T_{ij}) : 1 \leq i \leq [n/2], 1 \leq j \leq m\}$ and $\{(X_{ij}, T_{ij}) : [n/2] + 1 \leq i \leq n, 1 \leq j \leq m\}$. The first subset is utilized for mean estimation, while the second serves for covariance estimation. For the sake of simplicity, the notations indexing these two subsets are omitted in the subsequent discussions, provided that doing so does not lead to any ambiguity.*

### 3.4. Robust Loss Function Selection

In this section, we discuss the selection of the hyper-parameter $\kappa$ for the locally smooth robust loss function $\rho_1$ in (8). The choice of $\kappa$ plays a significant role in shaping the behavior of $\rho_1$, which in turn affects the definition and estimation of the robust mean $\mu_r$. It is essential to recognize that the bandwidth parameters, $h_\mu$ and $h_C$, play a distinct role: although they contribute to the estimation process, their influence on the true value of the robust mean is absent. Furthermore, as the hyper-parameter $\kappa$ approaches zero, the robust mean $\mu_r$ converges to the median. Consequently, the effect of hyper-parameter $\kappa$ on the estimation of $\mu_r$ becomes negligible for small values of $\kappa$.

Based on the above considerations, we select the hyper-parameter $\kappa$ by evaluating its performance in robust mean estimation. Specifically, $\kappa$ is chosen through a $K$-fold cross-validation procedure, in which entire subjects are left out. This process involves randomly and equally dividing $n$ subjects into $K$ subsets $\{I_1, \ldots, I_K\}$ and using each subset for validation in turn, aiming to minimize the empirical mean square error

$$\sum_{k=1}^{K} \sum_{i \in I_k} \sum_j \left( X_{ij} - \hat{\mu}_r^{(-k)}(T_{ij}) \right)^2.$$

Here, $\hat{\mu}_r^{(-k)}$ represents the robust mean estimate obtained using the training set after excluding the subjects in $I_k$. Once the hyper-parameter $\kappa$ and the corresponding locally smooth robust loss function $\rho_1$ are determined, the subsequent robust mean, covariance, and eigenfunction estimates are computed based on this tuned $\kappa$.

## 4. Theory

**Mean Estimation.** We initially examine the asymptotic property of the robust mean estimation in (11) under the standard regularity conditions. These conditions are provided as follows.

Assumption 4.1.

(1) *The kernel function $K$ is a symmetric, smooth probability density function, possessing compact support set on the interval $[-1, 1]$. The moment of this function is given by $\|K\|_{l_1, l_2} = \int K^{l_1}(u) u^{l_2} du$.*

(2) *The third-order mixed derivative $\frac{\partial^3}{\partial \beta \partial t^2} F(\beta, t)$ is bounded for the function $F(\beta, t) = \mathsf{E} \rho \left( X(t) - \beta \right)$.*

*(3) The rescaling function ψ is bounded.*

*(4) The robust mean $\mu_r(t)$ exists and is unique. Moreover, there is a positive constant $C_1 > 0$ such that for all t and for β near $\mu_r(t)$*

$$F(\beta, t) - F(\mu_r(t), t) - C_1'(\beta - \mu_r(t))^2 \geq 0. \tag{14}$$

Assumption (1) is frequently employed in kernel regression, while (2) assumes regularity on the underlying population $X$ through $F(\beta, t)$. Condition (3) necessitates that the robust loss function ρ exhibits growth no more rapid than the absolute loss. Simultaneously, (4) corresponds to the convexity of $F(\cdot, t)$ around $\mu_r(t)$. We examine various smoothed robust loss functions in Section 2.3. These robust losses conform to Assumptions (3)(4), as corroborated by Lemma 4.1.

Lemma 4.1. *Assumptions 4.1(3)(4) are satisfied for the robust loss functions $\rho_2$, $\rho_3$ (9), as well as for $\rho_1$ (8) given that the support set of the density function of $X(t)$ is connected.*

Recalling Assumption 3.1 on the random design, we obtain pointwise, locally uniform and $L^2$ convergence rates for the robust mean estimate (11), which aligns with the classic result presented in (Zhang and Wang, 2016, Theorem 4.1).

Theorem 4.1. *Under Assumptions 3.1 and 4.1, if $h_\mu \to 0$ and $nmh_\mu \to \infty$, then for any $t \in \mathcal{T}$ and $h = O(h_\mu)$, the pointwise, local uniform and $L^2$ convergence rates of the robust mean estimate given in (11) satisfy*

$$\max\left\{|\hat{\mu}_r(t) - \mu_r(t)|, \sup_{\tau \in B(t;h)} |\hat{\mu}_r(\tau) - \mu_r(\tau)|, \|\hat{\mu}_r - \mu_r\|_{L^2[0,1]}\right\} = O_p\left\{h_\mu^2 + \sqrt{\frac{1}{n} + \frac{1}{nmh_\mu}}\right\}.$$

We remark that both the local uniform and $L^2$ convergence properties are necessary for obtaining the convergence rates of the estimated covariance and eigenfunctions coupled with the robust mean estimate. It is important to note that the local uniform rate coincides with the pointwise rate but is distinct from the global uniform convergence rate, which entails an additional compensation factor $\log n$, as indicated by Zhang and Wang (2016, Theorem 5.1). The underlying cause of this phenomenon is that $\mathbb{E}\{K_{h_\mu}(T - t)\} = 1$ for a fixed point $t$, whereas $\mathbb{E}\{\sup_{t \in \mathcal{T}} K_{h_\mu}(T - t)\} = 1/h_\mu \to \infty$. Consequently, an extra $\log n$ factor is required to counterbalance this divergence in the context of global uniform convergence. For local uniform convergence, if it holds that $h = O(h_\mu)$ and, therefore, $\mathbb{E}\{\sup_{\tau: |\tau - t| \leq h} K_{h_\mu}(T - \tau)\} = O(h/h_\mu) = O(1)$, no compensation is needed.

The following corollary demonstrates the phase transition phenomenon at $m \asymp n^{1/4}$ in robust mean estimation, which is consistent with the classical result presented in (Zhang and Wang, 2016, Corollary 4.2). By selecting an appropriate value for $h_\mu$, it is shown that as long as $m$ grows at a rate not less than $n^{1/4}$, the convergence rate remains unaffected at the parametric order, specifically, $n^{-1/2}$. Conversely, if $m$ does not grow at this rate, its influence on the convergence rate is observed, resulting in the nonparametric rate of convergence.

Corollary 4.1. *Assume the conditions of Theorem 4.1 hold.*

- *When $m \gtrsim n^{1/4}$ and $h_\mu \asymp n^{-1/4}$ and*

$$\max\left\{|\hat{\mu}_r(t) - \mu_r(t)|, \sup_{\tau \in B(t;h)} |\hat{\mu}_r(\tau) - \mu_r(\tau)|, \|\hat{\mu}_r - \mu_r\|_{L^2[0,1]}\right\} = O_p\left(n^{-1/2}\right).$$

- *When $m \lesssim n^{1/4}$, the proper bandwidth is $h_\mu \asymp (nm)^{-1/5}$ and*

$$\max\left\{|\hat{\mu}_r(t) - \mu_r(t)|, \sup_{\tau \in B(t;h)} |\hat{\mu}_r(\tau) - \mu_r(\tau)|, \|\hat{\mu}_r - \mu_r\|_{L^2[0,1]}\right\} = O_p\left((nm)^{-2/5}\right).$$

Moreover, we establish the asymptotic normality property of the robust mean estimate in the following theorem, which again is in accordance with the classical result presented by Zhang and Wang (2016, Theorem 3.1).

Theorem 4.2. *Under Assumptions 3.1 and 4.1, if $h_\mu \to 0$ and $nmh_\mu \to \infty$, then for any $t \in \mathsf{T}$, the asymptotic normality of robust mean estimate given in (11) satisfy*

$$AV^{-1/2}\left(\hat{\mu}_r(t) - \mu_r(t) - AB\right) \xrightarrow{D} \mathrm{Normal}(0, 1)$$

*where the asymptotic variance $AV$ and asymptotic bias $AB$ are given by*

$$AV = \frac{1}{nmh_\mu}\|K\|_{2,0}^2 \frac{1}{f_T(t)} + \frac{m-1}{n}\mathsf{E}\left\{\psi^{2'}(X(t) - \mu_r(t))\left(\frac{\partial^2}{\partial \beta^2} F(\mu_r(t), t)\right)^{-2}\right\};$$

$$AB = -\frac{1}{2} h_\mu^2 \|K\|_{1,2} \frac{\partial^3}{\partial \beta \partial^2 t} F(\mu_r(t), t) \left(\frac{\partial}{\partial^2 \beta} F(\mu_r(t), t)\right)^{-1}.$$

**Covariance Estimation.** We next establish the convergence rate of the robust covariance estimate in (12).

Theorem 4.3. *Under Assumptions 3.1 and 4.1, if $h_\mu, h_C \to 0$, $nmh_\mu, nm^2 h_C^2 \to \infty$, and the robust functional covariance $C_r(s,t)$ in (4) is twice continuously differentiable, then for any $s, t \in \mathsf{T}$ and $h_C = O(h_\mu)$, the covariance estimate in (12) satisfies*

$$|\hat{C}_r(s,t) - C_r(s,t)| = O_p\left(h_\mu^2 + h_C^2 + \sqrt{\frac{1}{n} + \frac{1}{nmh_\mu} + \frac{1}{nm^2 h_C^2}}\right).$$

Note that the condition $h_C = O(h_\mu)$ is essential for leveraging the local uniform convergence rate of the robust mean estimate, as presented in Theorem 4.1. By suitably selecting bandwidths $h_\mu$ and $h_C$ in accordance with Corollary 4.2, it is possible to achieve convergence rates that are in line with the well-established classic rates (Zhang and Wang, 2016, Corollary 4.4). Furthermore, we once again witness a phase transition phenomenon occurring at $m \asymp n^{1/4}$, which can be described as follows.

Corollary 4.2. *Assume the conditions of Theorem 4.3.*

- *When $m \gtrsim n^{1/4}$, the proper bandwidths are $h_\mu \asymp h_C \asymp n^{-1/4}$ and*

$$|\hat{C}_r(s,t) - C_r(s,t)| = O_p\left(n^{-1/2}\right).$$

- When $m \lesssim n^{1/4}$, the proper bandwidths are $h_\mu \asymp h_C \asymp n^{-1/6} m^{-1/3}$ and

$$\|\hat{C}_r(s,t) - C_r(s,t)\| = O_p\left(n^{-1/3} m^{-2/3}\right).$$

In addition to the similarities with the mean estimation case, we observe that specifically for $m \lesssim n^{1/4}$, the choice of $h_\mu \asymp n^{-1/6} m^{-1/3}$ is necessary to satisfy $h_C = O(h_\mu)$. This selection is larger than the one in Corollary 4.1, where $h_\mu \asymp (nm)^{-1/5}$ for $m \lesssim n^{1/4}$. This observation suggests that in order to achieve the optimal convergence rate for covariance estimate, over-smoothing might occur in the robust mean estimation process.

**Eigenfunction Estimation.** Indeed, the perturbation bound for a diverging series of estimated eigenfunctions derived from discretely observed data remains an unresolved issue, despite the established results for fully observed functions presented in Hall and Hosseini-Nasab (2006) and Hall and Horowitz (2007), and the result for a fixed number of eigenfunctions in Hall et al. (2006). Recently, an advanced perturbation result was introduced in Zhou et al. (2022), which enhances the existing theoretical foundations for eigenfunction estimates based on kernel-type (non-robust) covariance estimation. In this study, we utilize these novel techniques to determine the convergence rates for a diverging number of robust eigenfunction estimates in (13). The standard regularity conditions are summarized as follows.

Assumption 4.2.

*(1) The observed time points $\{T_{ij}\}$ are independent and uniformly distributed on $[-\delta_0, 1+\delta_0]$ with a small extension parameter $\delta_0 \geq h_C$.*

*(2) The rescaling function $\psi$ is Lipschitz and the robust covariance $C_r$ has bounded forth derivatives.*

*(3) The FPC scores have finite forth moments and satisfy $\mathsf{E}\,\xi_k^4 \lesssim \lambda_k^2$.*

*(4) The underlying eigenvalues admit a polynomial decay $\lambda_k \asymp k^{-a}$ and the spacing is $\lambda_k - \lambda_{k+1} \asymp k^{-a-1}$ for a constant $a > 1$.*

*(5) The derivatives of the eigenfunction are uniformly bounded in the sense that*

$$\sup_{0 \leq r \leq \lceil b \rceil + 1} \left| \frac{d^r}{dt^r} \varphi_k(t) \right| \lesssim k^{rb/2},$$

*for a positive constant $b > 0$ satisfying $2b + 1 < 2a$.*

Assumption (1) stipulates that the observed time points are uniformly distributed over the interval $[-\delta_0, 1 + \delta_0]$, where $\delta_0$ is a small extension parameter on the same scale as the bandwidth $h_C$. This technical assumption is made to simplify derivations and notations (Hall, 1984; Zhou et al., 2022). The regularity conditions for moments (2) and (3), as well as the polynomial decay of eigenvalues (4), are also commonly employed in functional data analysis (Yao et al., 2005; Hall and Horowitz, 2007; Dou et al., 2012). Condition (5) characterizes the frequency increment of eigenfunctions through the bound of their derivatives. The requirement $2b + 1 < 2a$ is necessary to control the inflated bounds of these derivatives due to decreasing eigenvalues, and to characterize the convergence of an increasing number of estimated eigenfunctions (Zhou et al., 2022). For example, standard orthonormal bases such as Fourier, Legendre, and wavelet bases satisfy the condition $b = 2$.

**Theorem 4.4.** *Under Assumptions 4.1 and 4.2, the estimated eigenfunctions (13) for $k \leq k_0$ with $h_C^4 k^{2a+2b} = O(1)$ satisfy*

$$\|\hat{\varphi}_k - \varphi_k\|_{L^2} = O_p\left(k^{a/2+1}h_\mu^2 + \sqrt{\frac{k^{a+2}}{n} + \frac{k^{a+2}}{nmh_\mu}} + k^{b+1}h_C^2 + \sqrt{\frac{k^2}{n}\left(1 + \frac{k^{2a}}{m^2}\right) + \frac{k^a}{nmh_C}\left(1 + \frac{k^a}{m}\right)}\right).$$

*Moreover, if the robust mean $\mu_r$ is known a priori, one has*

$$\|\hat{\varphi}_k - \varphi_k\|_{L^2} = O_p\left(k^{b+1}h_C^2 + \sqrt{\frac{k^2}{n}\left(1 + \frac{k^{2a}}{m^2}\right) + \frac{k^a}{nmh_C}\left(1 + \frac{k^a}{m}\right)}\right).$$

The primary message conveyed by this novel perturbation result is that the number of eigenfunctions, represented by $k_0$, that can be well estimated permits growth alongside the sample size $n$ (jointly determined with the sampling frequency $m$, as the bandwidth $h_C$ is determined by $n$ and $m$), and is also influenced by the regularity parameters $a$ and $b$. Specifically, the eigenvalue spacing, characterized by the decay rate $a$, delineates the degree of separation between adjacent eigenvalues, subsequently impacting the estimation accuracy attainable for the eigenfunctions.

This result deviates from the perturbation bounds established for non-robust cases as presented in the work of Zhou et al. (2022). Here, we incorporate the unknown mean into our eigenfunction estimates; however, addressing the robust mean estimate, which lacks an analytic expression, poses a significant technical challenge. Nevertheless, this result serves as a fundamental basis for investigating more advanced models reliant on robust FPCA of discretely observed data, including functional linear regression, among others. To elucidate the implications of this newly-developed perturbation bound, we carefully select appropriate bandwidths $h_\mu$ and $h_C$, subsequently revealing the interrelation among the truncation parameter $k_0$, the sample size $n$, and the sampling frequency $m$ in the following corollary.

**Corollary 4.3.** *Assume the conditions of Theorem 4.4. The bandwidths are selected as $h_\mu \asymp (nm)^{-1/5}$ and $h_C \asymp k_0^{(a-2b-2)/5}(nm)^{-1/5}(1 + k_0^a/m)^{1/5}$.*

- *When $m \gtrsim (1 + k_0^{a+b/2-2}(1 + k_0^a/m)^{-3/2})n^{1/4}$, the convergence rate is*

$$\|\hat{\varphi}_{k_0} - \varphi_{k_0}\|_{L^2} = O_p\left(k_0^{a/2+1}n^{-1/2} + (1 + k_0^a/m)k_0n^{-1/2}\right).$$

*Furthermore, if $m \gtrsim k_0^a$, the rate becomes $O_p\left(k_0^{a/2+1}n^{-1/2}\right)$.*

- *When $m \lesssim n^{1/4}$ and $m \lesssim k_0^{a+b/2-2}(1 + k_0^a/m)^{-3/2}n^{1/4}$, the convergence rate is*

$$\|\hat{\varphi}_{k_0} - \varphi_{k_0}\|_{L^2} = O_p\left(k_0^{a/2+1}(nm)^{-2/5} + (1 + k_0^a/m)^{2/5}k_0^{(2a+b+1)/5}(nm)^{-2/5}\right).$$

*Furthermore, if $m \lesssim k_0^a$, the rate becomes $O_p\left(k_0^{a/2+1}(nm)^{-2/5} + k_0^{(4a+b+1)/5}n^{-2/5}m^{-4/5}\right).$*

*If the robust mean $\mu_r$ is known, the bandwidth $h_C \asymp k_0^{(a-2b-2)/5}(nm)^{-1/5}(1 + k_0^a/m)^{1/5}$.*

- When $m \gtrsim k_0^{a+b/2-2}(1 + k_0^a/m)^{-3/2} n^{1/4}$, the convergence rate is

$$\|\hat{\varphi}_{k_0} - \varphi_{k_0}\|_{L^2} = O_p\left((1 + k_0^a/m) k_0 n^{-1/2}\right).$$

Furthermore, if $m \gtrsim k_0^a$, the rate becomes $O_p\left(k_0 n^{-1/2}\right)$.

- When $m \lesssim k_0^{a+b/2-2}(1 + k_0^a/m)^{-3/2} n^{1/4}$, the convergence rate is

$$\|\hat{\varphi}_{k_0} - \varphi_{k_0}\|_{L^2} = O_p\left((1 + k_0^a/m)^{2/5} k_0^{(2a+b+1)/5} (nm)^{-2/5}\right).$$

Furthermore, if $m \lesssim k_0^a$, the rate becomes $O_p\left(k_0^{(4a+b+1)/5} n^{-2/5} m^{-4/5}\right)$.

The corollary characterizes an intriguing phase transition phenomenon in the convergence of robust eigenfunction estimation. For illustration, we consider $\hat{\varphi}_{k_0}$ as estimated eigenfunctions $\hat{\varphi}_k$ for $1 \leq k < k_0$ exhibit faster convergence. We observe that the sampling frequency $m$ is partitioned into three magnitude orders in relation to the sample size $n$ and the truncation $k_0$, which are expressed as $m \asymp n^{1/4}$, $m \asymp k_0^{a+b/2-2}(1 + k_0^a/m)^{-3/2} n^{1/4}$ and $m \asymp k_0^a$. The first and second transition orders are attributed to the mean and covariance estimates, respectively. Nonetheless, this differs from Corollary 4.2, where these two estimates share the same transition order $m \asymp n^{1/4}$. Notably, the transition order $m \asymp k_0^{a+b/2-2}(1 + k_0^a/m)^{-3/2} n^{1/4}$ highlights the increased complexity of eigenfunction estimation compared to mean/covariance estimation, as diminishing gaps between eigenvalues make the estimation of high-order eigenfunctions more challenging. The order $m \asymp k_0^a$ arises due to the additional term $k_0^a/m$ presented in the eigenfunction estimation, which amplifies the rate when the sampling frequency $m$ is smaller than $k_0^a$ (i.e., the data becomes sparse).

We would also like to highlight that the results obtained above are consistent with existing theory in the special case. For instance, when the number of eigenfunctions and sampling frequency are bounded, i.e., $k_0$ and $m$ are fixed corresponding to the "very sparse" (or longitudinal) data, the convergence takes the form $h_\mu^2 + h_C^2 + (nh_\mu)^{-1/2} + (nh_C)^{-1/2}$. This is the typical one-dimensional nonparametric rate, which aligns with the findings in Hall et al. (2006). Conversely, if the sampling frequency $m$ exceeds these three magnitude orders, the convergence attains an optimal parametric rate of $k_0/\sqrt{n}$. This rate cannot be further improved as $m$ increases and is akin to the case where random functions are fully observed (Hall and Horowitz, 2007; Wahl, 2022).

We conclude this section by emphasizing that classic perturbation theory (Bosq, 2000) is unable to provide the improved rates presented in this section, as it only produces a rough bound for $1 \leq k \leq n$.

$$\|\hat{\varphi}_k - \varphi_k\|_{L^2} = O_p\left(k^{a+1}\left\{h_\mu^2 + h_C^2 + \left[\frac{1}{n} + \frac{1}{nmh_\mu} + \frac{1}{nm^2 h_C^2}\right]^{1/2}\right\}\right), \quad (15)$$

which is directly inherited from the covariance estimation in Theorem 4.3. This rough bound can be seen to be sub-optimal in two different aspects. Firstly, the term $h_C^2 + n^{-1/2} h_C^{-1}$ only approaches the two-dimensional nonparametric rate when $k$ and $m$ are bounded. This is in contrast to the expected behavior of the eigenfunction estimates, which enjoy the one-dimensional nonparametric rate $h_C^2 + (nh_C)^{-1/2}$ as demonstrated by Hall et al. (2006). Secondly, when the observed times become dense, this rate changes to $k^{a+1}/\sqrt{n}$, which falls short of attaining the optimal rate, $k/\sqrt{n}$, as observed in fully observed cases (Wahl, 2022).

## 5. Simulation

As discussed earlier, the existing works on robust functional data modeling have come in different definitions or forms of the mean, covariance and eigen-analysis, which cannot be consolidated into a unified or comparable framework. Consequently, our primary focus lies in assessing the quality of robust estimation provided by our proposed robust mean and covariance estimators, and evaluating their consistency. The results of robust score estimation are deferred to the Supplementary Material for space economy. We also discuss a comparison with the Robust Smoothing Spline (RSP) method that investigates robust mean estimation (Kalogridis and Van Aelst, 2022).

In our assessment, we consider observations that follow a Gaussian distribution as benchmark, as well as those with heavy-tailed or contaminated and skewed distributions. The underlying population is characterized by $X = \sum_{k=1}^{\infty} \xi_k \phi_k$ on $T = [0, 1]$, where the underlying functions are $\phi_k = \sqrt{2}\sin(k\pi t)$, and the distributions of the scores $\{\xi_k\}_k$ are as follows.

(1) Normal distribution $\xi_k \sim \text{Normal}(0, k^2)$, serving as the benchmark;

(2) $t$-distribution $\xi_k \sim t_k$, which implying the absence of classic mean and covariance;

(3) Symmetric log-normal distribution $\xi_k \sim \text{SLN}(0, k^2)$. In this case, $X$ exhibits heavy tails; however, its classic mean and covariance still exist.

We also examine the case of centralized-Beta distributed scores ‖ in which the robust mean $\mu_r$ does not equal the classic mean $\mu_c = 0$ when employing non-square loss functions.

(4) Centralized-Beta$(2k, k)$ that is asymmetrically distributed.

To demonstrate the impact of contamination, each observation from the centralized-Beta distribution is replaced by a sample from Normal$(10, 0.1^2)$ with probabilities $\alpha\% = 10\%, 20\%$. In this construction, both the symmetric log-normal and $t$ distributions exhibit heavy-tailed behavior to varying extents, while the centralized-Beta represents skewed and contaminated scenarios. In comparison, the normal distribution serves as a benchmark. The subsequent analysis considers the following robust loss functions.

- The squared loss function $\rho_0(x) = x^2$, representing the classic case where no robust approach is utilized;

¶A random variable $Z$ is said to follow a symmetric log-normal distribution, denoted by SLN$(\mu, \sigma^2)$, if $Z = Z_1 Z_2$, where $Z_1$ and $Z_2$ are independent, $Z_1$ follows a log-normal distribution LN$(\mu, \sigma^2)$, and $Z_2$ follows a binary distribution with $P(Z_2 = 1) = P(Z_2 = -1) = 0.5$.

‖The centralized-Beta$(\alpha, \beta)$ distribution refers to $Z = Z_1 - \mathsf{E}(Z_1)$, where $Z_1 \sim \text{Beta}(\alpha, \beta)$.

- The locally smooth robust loss function $\rho_1(x) = |x|I_{|x|>\kappa} + \frac{1}{8\kappa}(-x^4 + 6\kappa^2 x^2 + 3\kappa^4)I_{|x|\leq\kappa}$ in (8), utilized with predetermined hyper-parameter values of $\kappa = 10^{-3}$, $10^{-2}$, $10^{-1}$, 1, and a tuned $\kappa$ following the two-fold cross-validation approach detailed in Section 3.4.

- The globally smooth robust loss functions $\rho_2(x) = \log(\cosh(x))$ and $\rho_3(x) = 2\pi^{-1}x\arctan(x) - \pi^{-1}\ln(1+x^2)$ in (9).

As the analytical expressions for the robust mean $\mu_r$ and covariance $C_r$ are typically unavailable, we resort to computing their estimates by utilizing a large number of independent samples. Specifically, we compute $\mu_r(t)$ and $C_r(s,t)$ for every $s, t \in \mathcal{T}$ using a significant number ($10^6$) of samples, and repeat this procedure 100 times to obtain Monte Carlo averages for reliable approximation and evaluation of their true values.

To accommodate the discretely observed functional data, we generate $n = 100, 200$ sample curves that are observed on $m = 5, 10$ time points $T_{ij}$ drawn from Uniform(0, 1). The local regression method employs the compactly supported polynomial kernel $K(u) = \frac{70}{8}(1-u^2)^3$, with the bandwidths $h_\mu$ and $h_C$ selected by minimizing the two-fold cross-validation errors. Tables 1 and 3 display the (relative) integral mean squared errors of the mean and covariance estimates, respectively, which are derived from the average of 100 Monte Carlo runs. These findings indicate that our proposed robust mean and covariance estimates significantly reduce the estimation errors for non-Gaussian distributions in comparison with the non-robust method (the squared error $\rho_0$), and they also converge towards their true values as the sample size $n$ and sampling rate $m$ increase. The performance of locally and globally smooth loss functions exhibits minor disparities across diverse score distributions. Notably, $\rho_1$ demonstrates superior performances in contaminated cases, particularly when the contamination level reaches 20% in both tables. The outcomes of $\rho_1$ with the tuned $\kappa$ are on par with the minimum errors yielded by the predetermined $\kappa$ candidates, implying that the adopted tuning approach effectively optimizes the robust mean estimation performance, as anticipated.

We conclude this section by comparing our proposed method to the RSP (Robust Smoothing Spline) method that examines robust mean estimation for discretely observed functional data (Kalogridis and Van Aelst, 2022). The underlying settings in our comparison are similar to those above, and slight modifications are made to the centralized-Beta($10k, k$) (multiplied by 10) scores to emphasize the asymmetry. We generate $n = 100$ samples $\{X_i\}_{i=1}^n$, where each curve $X_i$ is observed at $m = 5$ uniformly distributed time points $\{T_{ij}\}_{j=1}^m$, for illustration purposes. Table 2 presents the integral mean square errors using the Monte Carlo method over 100 independent runs. It reveals that our method outperforms the RSP method in contaminated and skewed cases, but exhibits the opposite behavior in heavy-tail cases.

## 6. Real Data Analysis

We utilize the proposed framework to examine longitudinal diffusion tensors obtained from the Alzheimer's Disease Neuroimaging Initiative (ADNI) database. The principal objective of ADNI is to investigate whether a combination of serial magnetic resonance

**Table 1.** Shown are the integral mean squared errors for the mean estimation, $\int E\left(\hat{\mu}_r(t) - \mu_r(t)\right)^2 dt$. These values are calculated using 100 Monte Carlo simulations, and the corresponding standard errors are provided in parentheses. Here, SLN denotes the symmetric log-normal distribution, while Beta, $\alpha$% refers to the centralized-Beta distribution with $\alpha$% contamination.

| $\rho$ | n | m | Normal | t | SLN | Beta,10% | Beta,20% |
|---|---|---|---|---|---|---|---|
| $\rho_0$ | 100 | 5 | .0493(.0043) | 12.15(2.107) | 9.818(1.615) | 1.110(.0276) | 4.060(.0751) |
| | | 10 | .0375(.0039) | 5.805(1.367) | 5.939(1.264) | 0.990(.0195) | 4.097(.0542) |
| | 200 | 5 | .0229(.0017) | 12.00(2.508) | 6.605(1.184) | 1.041(.0207) | 4.044(.0557) |
| | | 10 | .0187(.0016) | 9.370(2.009) | 3.894(0.884) | 1.009(.0156) | 3.999(.0400) |
| $\rho_1$ $\kappa = 10^{-3}$ | 100 | 5 | .0499(.0057) | .0436(.0050) | .0641(.0056) | .0021(.0002) | .0073(.0001) |
| | | 10 | .0419(.0048) | .0450(.0062) | .0570(.0056) | .0018(.0001) | .0072(.0001) |
| | 200 | 5 | .0221(.0031) | .0270(.0032) | .0346(.0039) | .0015(.0001) | .0070(.0001) |
| | | 10 | .0147(.0018) | .0177(.0022) | .0334(.0033) | .0013(.0001) | .0069(.0001) |
| $\rho_1$ $\kappa = 10^{-2}$ | 100 | 5 | .0441(.0048) | .0531(.0064) | .0716(.0073) | .0023(.0003) | .0081(.0006) |
| | | 10 | .0372(.0047) | .0428(.0055) | .0773(.0111) | .0016(.0001) | .0083(.0005) |
| | 200 | 5 | .0196(.0022) | .0230(.0028) | .0281(.0032) | .0015(.0001) | .0079(.0004) |
| | | 10 | .0172(.0025) | .0212(.0024) | .0350(.0037) | .0013(.0001) | .0068(.0003) |
| $\rho_1$ $\kappa = 10^{-1}$ | 100 | 5 | .0423(.0052) | .0395(.0036) | .0706(.0064) | .0024(.0002) | .0074(.0004) |
| | | 10 | .0366(.0064) | .0439(.0056) | .0616(.0075) | .0020(.0001) | .0073(.0004) |
| | 200 | 5 | .0207(.0024) | .0332(.0041) | .0359(.0039) | .0016(.0001) | .0077(.0004) |
| | | 10 | .0206(.0026) | .0224(.0025) | .0335(.0036) | .0015(.0001) | .0077(.0003) |
| $\rho_1$ $\kappa = 1$ | 100 | 5 | .0413(.0046) | .0480(.0043) | .0770(.0088) | .0079(.0004) | .0358(.0011) |
| | | 10 | .0369(.0039) | .0538(.0051) | .0538(.0057) | .0078(.0004) | .0365(.0010) |
| | 200 | 5 | .0235(.0026) | .0292(.0031) | .0472(.0045) | .0071(.0002) | .0351(.0008) |
| | | 10 | .0209(.0035) | .0224(.0026) | .0297(.0034) | .0069(.0002) | .0346(.0006) |
| $\rho_1$ tuned $\kappa$ | 100 | 5 | .0466(.0057) | .0594(.0082) | .0710(.0081) | .0022(.0002) | .0082(.0004) |
| | | 10 | .0445(.0057) | .0415(.0047) | .0625(.0072) | .0021(.0002) | .0075(.0004) |
| | 200 | 5 | .0276(.0033) | .0245(.0028) | .0367(.0036) | .0019(.0001) | .0073(.0004) |
| | | 10 | .0208(.0032) | .0168(.0018) | .0264(.0034) | .0015(.0001) | .0071(.0004) |
| $\rho_2$ | 100 | 5 | .0518(.0047) | .0562(.0056) | .0682(.0080) | .0177(.0009) | .0779(.0023) |
| | | 10 | .0349(.0037) | .0495(.0056) | .0640(.0065) | .0146(.0006) | .0775(.0016) |
| | 200 | 5 | .0217(.0028) | .0296(.0033) | .0372(.0039) | .0161(.0005) | .0754(.0015) |
| | | 10 | .0177(.0022) | .0251(.0027) | .0325(.0032) | .0148(.0003) | .0762(.0012) |
| $\rho_3$ | 100 | 5 | .0434(.0041) | .0751(.0097) | .0856(.0076) | .0348(.0014) | .1941(.0178) |
| | | 10 | .0330(.0048) | .0488(.0049) | .0815(.0076) | .0330(.0011) | .1702(.0035) |
| | 200 | 5 | .0233(.0025) | .0344(.0031) | .0518(.0057) | .0336(.0008) | .1762(.0034) |
| | | 10 | .0198(.0023) | .0272(.0034) | .0330(.0036) | .0311(.0006) | .1674(.0025) |

**Table 2.** The integral mean square errors of our method and RSP method. These values are calculated using 100 Monte Carlo simulations, and the corresponding standard errors are provided in parentheses.

| | MSE | Normal | t | SLN | Beta,10% | Beta,20% |
|---|---|---|---|---|---|---|
| $\rho_1$ | Our Method | .0398(.0045) | .0445(.0042) | .0780(.0085) | .0259(.0020) | .0904(.0046) |
| | RSP | .0223(.0026) | .0226(.0028) | .0426(.0047) | .0719(.0042) | .1763(.0071) |
| $\rho_2$ | Our Method | .0470(.0049) | .0476(.0042) | .0731(.0090) | .0364(.0025) | .1482(.0064) |
| | RSP | .0384(.0052) | .0322(.0034) | .0417(.0045) | .0763(.0042) | .2302(.0085) |
| $\rho_3$ | Our Method | .0557(.0061) | .0605(.0064) | .0820(.0078) | .0622(.0037) | .2792(.0084) |
| | RSP | .0442(.0058) | .0418(.0046) | .0547(.0050) | .1008(.0048) | .3676(.0113) |

**Table 3.** Shown are the relative integral mean squared error for the covariance estimation, $\int E'\hat{C}_r(s,t) - G(s,t))^2 dsd \int C_r^2(s,t)dsdt$. These values are calculated using 100 Monte Carlo simulations, and the corresponding standard errors are provided in parentheses. Here, SLN denotes the symmetric log-normal distribution, while Beta, $\alpha\%$ refers to the centralized-Beta distribution with $\alpha\%$ contamination.

| $\rho$ | n | m | Normal | $t$ | SLN | Beta,10% | Beta,20% |
|---|---|---|---|---|---|---|---|
| $\rho_0$ | 100 | 5 | .1551(.0076) | .9999(.0001) | 17.25(10.97) | 150.9(15.58) | 440.8(31.28) |
| | | 10 | .1007(.0065) | 1.000(.0001) | 53.68(52.70) | 37.43(4.417) | 110.2(7.719) |
| | 200 | 5 | .0895(.0047) | .9984(.0011) | 39.46(38.43) | 92.52(9.638) | 288.1(26.15) |
| | | 10 | .0755(.0045) | .9997(.0001) | .9850(0.1538) | 22.49(1.725) | 67.06(5.922) |
| $\rho_1$ tuned $\kappa$ | 100 | 5 | .2965(.0134) | .2475(.0115) | .2789(.0112) | .2135(.0073) | .3122(.0080) |
| | | 10 | .2529(.0086) | .2183(.0077) | .2389(.0090) | .1485(.0042) | .2657(.0049) |
| | 200 | 5 | .2772(.0095) | .2284(.0082) | .2527(.0087) | .1535(.0052) | .2596(.0057) |
| | | 10 | .2453(.0070) | .2048(.0069) | .2251(.0064) | .1285(.0042) | .2392(.0038) |
| $\rho_2$ | 100 | 5 | .1210(.0040) | .1206(.0042) | .1512(.0060) | .3404(.0189) | .6587(.0307) |
| | | 10 | .0908(.0034) | .0754(.0030) | .1162(.0039) | .1894(.0080) | .4021(.0128) |
| | 200 | 5 | .0830(.0028) | .0723(.0025) | .1147(.0034) | .2249(.0087) | .5015(.0155) |
| | | 10 | .0685(.0016) | .0535(.0016) | .0877(.0022) | .1375(.0052) | .3511(.0096) |
| $\rho_3$ | 100 | 5 | .1055(.0035) | .1127(.0045) | .1761(.0073) | .5873(.0318) | 1.542(.1424) |
| | | 10 | .0791(.0029) | .0752(.0031) | .1099(.0042) | .2765(.0110) | .7761(.0287) |
| | 200 | 5 | .0738(.0024) | .0726(.0027) | .1073(.0043) | .3771(.0182) | .9609(.0469) |
| | | 10 | .0630(.0017) | .0472(.0014) | .0857(.0023) | .1971(.0077) | .5223(.0154) |

imaging, positron emission tomography, other biomarkers, and clinical and neuropsychological assessments can effectively measure the progression of mild cognitive impairment and early Alzheimer's disease (AD). For the most recent information, please visit www.adni-info.org.

The Fractional Anisotropy (FA) derived from longitudinal diffusion tensor imaging (a specialized type of diffusion-weighted magnetic resonance) characterizes the degree of anisotropy in water molecule diffusion within the hippocampus. Variations in FA values suggest that the diffusion of water molecules is constrained by structures, such as white matter fibers, which serve as pivotal markers for the potential detection of Alzheimer's disease. As a result, FA analysis holds significant implications for both clinical diagnostics (Lindberg et al., 2012) and scientific research (Pennec et al., 2006; Fletcher and Joshi, 2007; Shao et al., 2022) related to Alzheimer's disease.

In this study, we consider subjects who have at least three properly recorded FA observations. Consequently, the sample consists of $n = 136$ subjects with ages ranging from 65 to 85. Among them, $n_{CN} = 34$ subjects are cognitively normal (CN), whereas the remaining $n_{AD} = 102$ subjects (AD) developed mild cognitive impairment or Alzheimer's disease. On average, each subject has $\overline{m} = 5.7$ FA observations, indicating that the data is rather sparsely recorded. Figure 3 presents all observations of FA levels for the AZ group (left) and CN group (right), graphically illustrating the drastic distribution of the observations and emphasizing the necessity of robust modeling for the data analysis. We report the results obtained using the locally smooth robust loss function $\rho_1$ in (8) with tuned $\kappa = 1, 0.5995$ for AD and CN groups, respectively; the results for other loss functions $\rho_2$ and $\rho_3$ display similar patterns and are provided in the Supplementary

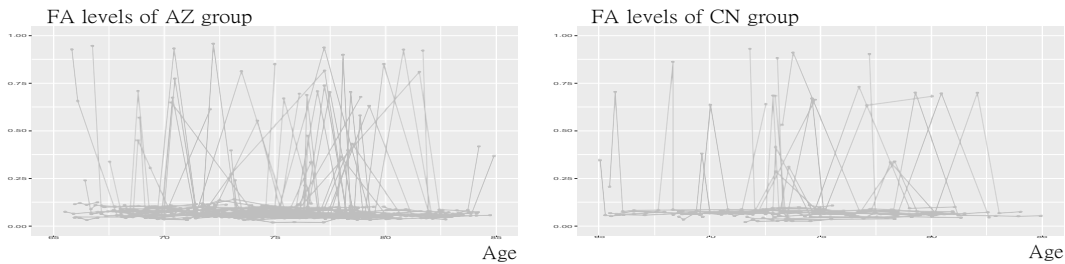

**Fig. 3.** The observations of FA levels in the AZ group (on the left) and the CN group (on the right).

Material.

The left panel of Fig. 4 displays the robust mean estimates of FA levels, where the bandwidths determined by two-fold cross-validation are $h_\mu = 12, 12$ for AD and CN groups, respectively. The mean trajectory of the AD group clearly exhibits a lower FA level within the majority range [68, 83]. This observation might suggest some damage to the hippocampal structure for the AD group. Additionally, the mean trajectories around the lower and upper ends, assigned as age 65 and 85, could result from the boundary effect of nonparametric smoothing. The middle panel of Fig. 4 presents the leading two robust eigenfunction estimates, accounting for (69.32%, 17.80%) and (64.44%, 16.06%) of the total variation for the AD and CN groups, respectively. When comparing these components side by side, they reveal different patterns between the two cohorts. For example, the first eigenfunction of the AD group is above-zero for the entire trajectory, which implies that individual FA trajectories in the AD group tend to deviate from their mean trajectory along the direction with below-normal FA levels. Conversely, the CN cohort exhibits a contrast between early and late ages, resulting in less pronounced below-normal FA patterns associated with AD. The right panel of Fig. 4 presents the leading two robust score estimates, suggesting that the scores of AD group tend to to cluster together, while CN group's incline to have a more scattered distribution. Above discussion suggests that the proposed robust method accurately captures distinct features of the AD and CN groups that align with previous research (Lindberg et al., 2012), even though the observed data are distributed drastically.

## Supplementary

The Supplementary Material includes simulations demonstrating the robust score estimates, additional results from real data applications, and the comprehensive proofs of the key theorems and related lemmas.

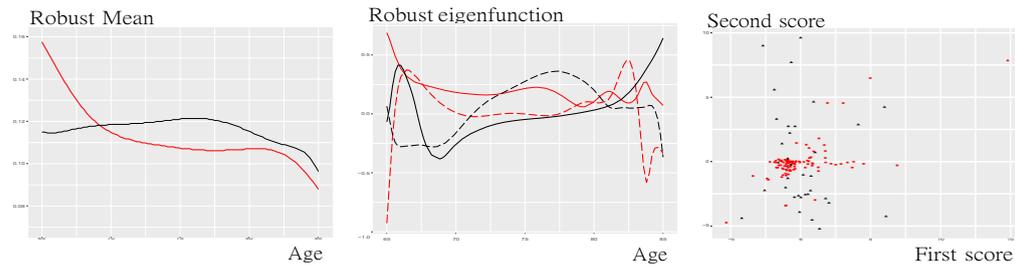

**Fig. 4.** The robust estimates of the mean, the leading two robust eigenfunctions and scores are displayed in the left, middle and right panels, respectively. The red legends indicate the AD group, while the black legends represent the CN group. In the middle panel, the solid and dashed curves correspond to the first and second robust eigenfunction estimates, respectively.